\documentclass[preprint2]{aastex}

\newcommand{\HII}{H\,{\sc ii}}

\newcommand{\um}{\,$\mu$m}
\newcommand{\kms}{\,km\,s$^{-1}$}

\slugcomment{Draft \today}

\shorttitle{[NeII] Kinematics in He 2-10  }
\shortauthors{Beck \% et al}

\begin{document}

\title{Ionized Gas Kinematics at High Resolution V: 
[NeII], Multiple Clusters, 
High Efficiency Star Formation and Blue Flows
 in He 2-10}

\author{Sara Beck\altaffilmark{1,2}, Jean Turner \altaffilmark{3}, John Lacy\altaffilmark{4}, Thomas Greathouse \altaffilmark{2,5}}
 \altaffiltext{1}{School of Physics and Astronomy, Tel Aviv University, Ramat Aviv ISRAEL 69978}
\altaffiltext{2}{Visiting Astronomer at the Infrared Telescope Facility, which is operated by the University of Hawaii under Cooperative Agreement no. NNX-08AE38A with the National Aeronautics and Space Administration, Science Mission Directorate, Planetary Astronomy Program.}
\altaffiltext{4}{Department of Astronomy, University of Texas at Austin, Austin Tx 78712}
\altaffiltext{3}{Department of Physics and Astronomy, UCLA, Los Angeles, CA 90095-1547}
\altaffiltext{5}{Southwest Research Institute, San Antonio Tx 78228-0510}

\begin{abstract}
We measured  the $12.8\mu$m [NeII] line in the dwarf starburst galaxy He 2-10 with the high-resolution spectrometer TeXeS on the NASA IRTF.  The data cube has diffraction-limited spatial resolution $\sim1\arcsec$ and total velocity resolution including thermal broadening of $\sim5$\kms~. This makes it possible to compare the kinematics of individual star-forming clumps and molecular clouds in the three dimensions of space and velocity, and allows us to determine star formation efficiencies. The kinematics of the ionized gas confirm that the starburst contains multiple dense clusters.  From the $M/R$ of the clusters and the $\simeq30-40$\% star formation efficiencies the clusters are likely to be  bound and long lived,  like globulars.  Non-gravitational features in the line profiles show how the ionized gas flows through the ambient molecular material, as well as a narrow velocity feature which we identify with the interface of the  \HII~region and a cold dense clump.  These data offer an unprecedented view of the interaction of embedded \HII~regions with their environment. 
 
  \end{abstract}


\section{Introduction}
High resolution images of nearby starbursts reveal that they often consist of massive clusters, containing hundreds or thousands of OB stars in volumes no more than a few pc across.  Massive stars are powerful sources of ionization, heat, and kinetic energy, and profoundly effect the environment and the development of the host galaxy.    How can so many clusters form in such small regions, in such short (few Myr) times?  Are star formation efficiencies higher in starbursts? How does feedback manifest itself on parsec scales? 

The youngest star clusters are deeply embedded and extinction to these regions is so high that that only radio and infrared can reliably 
 penetrate.   To understand the evolution of these embedded starbursts and their interactions with  their surroundings we need to know the gas kinematics. 
 We therefore use high-resolution mid-infrared spectroscopy to study the ionized gas kinematics of intense starbursts. 

In this paper, the  fifth in the series of  TEXES observations \citep{BLT10,BLT12,BTL13,2014ApJ...787...85B}, the target is a dwarf starburst galaxy.  He 2-10 is nearby, at 10.5 Mpc, and is undergoing an intense starburst which is believed to have been triggered by a minor accretion event 
\citep{KO95}.   The star forming region which has been mapped in the infrared and radio continuua \citep{BT01, RE11}  is seen to contain multiple embedded star clusters grouped into two large complexes, all within an extent of 5\arcsec\ or 225 pc. 
 He 2-10 is rich in molecular gas. CO emission in the star formation region was first mapped by \citet{KO95}.  More recent observations of CO (2-1)  with high spatial and spectral resolution by \citet{SA09} reveal numerous molecular clouds whose masses and densities are at the upper end of the range seen for Giant Molecular Clouds in our Galaxy. 

Our earlier papers observed the bulk motions of  ionized gas immersed in molecular material in galaxies with one or two massive star clusters  \citep{BLT10,BLT12}.  He 2-10 is more complicated, with many star clusters  in a complex molecular environment of multiple sub-clouds.  

  \citet{BKL97} measured the mid-infrared lines of [NeII], [SIV] and [ArIII] in He 2-10 with spectral resolution $\sim 30$\kms\ and did not detect any structure in the line profiles.  \citet{AH07} observed  $Br\alpha$ and $Br\gamma$ spectra in the star formation region with effective spectral resolution $~21$\kms\, set by the thermal width of HI. They  found that the profiles in many positions had a narrow core and a broad low-intensity feature on the blue side which they interpreted as weak outflow.   These results called for further study of the ionized gas kinematics with higher spectral resolution. \
  
We have accordingly observed the ionized gas via the [NeII] $12.8\mu$m line, a tracer which is almost extinction free and which permits maximum velocity resolution of ~5 \kms, 
allowing us to probe sub-sonic motions within and among the HII regions.
In the next sections we describe the observations and results 
and discuss what they reveal about the young massive clusters and the embedding clouds.

 \section{Observations and Data Reduction}

\subsection{Current Observations}
 He 2-10 was observed at the NASA Infrared Telescope Facility on Mauna Kea, Hawaii, on the night of 12 January 2012, with the TEXES spectrometer \citep{LA02}.  TEXES is an echelon grating spectrograph with a $256 \times 256 ~Si:As$ detector array.  It operates from $5-25$\um.  
 On the IRTF the TEXES beam is set by seeing and the diffraction limit to $\sim1.4\arcsec$.  Positions are found by offset pointing from visible stars and reproduce to better than 1\arcsec. 
 The observations  were taken in the high-resolution mode, with a plate scale~$0.36\arcsec\times0.9~\rm km~s^{-1}$ per pixel and spectral resolution $\sim3.5$\kms.  The 9.3 \arcsec ~long slit was scanned across the galaxy in steps of  0.7\arcsec~.  
 Atmospheric lines set the wavelength scale, and asteroids were flux calibration sources. We estimate the flux errors at $\pm0.0005~\rm erg(s~cm^{-1}~cm^2~sr)^{-1}$ and the absolute uncertainty at  $\approx$30\%. 
 
  The line was recorded in  0.9~\kms~steps, giving 4 points per resolution element.  The echelon order includes 256 points covering the velocity range from 723 to 963 \kms~ LSR. 
  The data do not go down to 0 level in the spectra, indicating that either the line has an extremely broad plateau, which has not been hinted at in any other observations, or we have detected the  continuum.  
  We  checked this by creating a merged spectrum including 8 orders of the echelon, covering a velocity range 100--1900 \kms\ in 3.6\kms\ bins, and established that there is a weak continuum of $0.0015\pm0.0005~\rm erg(s~cm^{-1}~cm^2~sr)^{-1}$  
across the entire range, consistent with the level found by Spitzer.  The presence of a continuum can result in distorted and misleading fits to line profiles.    We have therefore adjusted the original data by subtracting off the continuum, and it is this subtracted data that we analyse and fit in  the rest of the paper. 

The atmosphere in the wavelength region scanned is mostly transparent.  Transmission at Mauna Kea \footnotemark~is around 95\% or more over most of the wavelength region for  [NeII] at velocities between 620 and 1120 \kms~. There are several weak ($<10\%$ deep) atmospheric wiggles throughout the range, which increase the noise.   The grating efficiency is not uniform across the scan but degrades as the wavelength gets further from the blaze angle. This results in non-uniform noise which increases towards the blue side of the line peak, and at velocities lower than ~840\kms~ is $\approx30$\% higher than on the red side of the line. This will limit our efforts to fit the lines, as will be discussed below. 
\footnotetext{http://www.gemini.edu/?q=node/10789}

 \section{Results}
\subsection{Integrated Spatial Distribution}
 Figure 1 shows the spatial distribution of the integrated [NeII] line emission, obtained by collapsing the TEXES data cube along the velocity axis, as well as the $12\mu$m continuum  and 2 cm radio emission. The spatial distribution of the [NeII] line agrees well with that of the mid-infrared and radio continuum emission. Both display two sources offset mostly E-W, at $P.A.\sim105^{\circ}$.  They extend $\sim5-6\arcsec$ E-W and $\sim2-3\arcsec$ N-S, and the eastern source is the stronger.  In the radio and infrared continua there is also a compact source between the main sources. It has a non-thermal radio spectrum and is coincident with a hard X-ray source; \citet{RE11} believe it to be  a mini-AGN with an accreting black hole.  It does not appear in the [NeII] map, consistent with \citet{RE11}: AGNs create high-excitation lines like [NeV], while [NeII] is associated with the softer radiation fields of star formation regions \citep{PE11}.

\subsection{[NeII] Flux and $Ne^+$ Abundance}
The total [NeII] flux summed over the full line extent of 150\kms~ and the entire region is $4.35\times10^{-12}~\rm erg~s^{-1}~cm^{-2}$, with absolute uncertainty $\approx$30\%.  The Spitzer IRS SH module with a $4-5\arcsec$ beam measured $10\pm0.05$ Jy, which is $3.8\pm0.002\times10^{-12}~\rm erg~s^{-1}~cm^{-2}$ [NeII] flux.  Within the considerable uncertainties that arise in converting Jy to flux for an unresolved line, this is consistent with TEXES having recovered the entire Spitzer flux.   The strength of the neon line is related to the thermal radio continuum as both depend on the emission measure, $EM= n_e^2l ~\rm (cm^{-6}~pc)$.   He 2-10 has (unusually for a dwarf galaxy) solar or near solar metallicity, and for the solar neon abundance of $8.3\times10^{-5}~n(H)$ and the collision strengths of \citet{OF06}, $F([NeII])(cgs)=2.0\times10^{-10}F_{5 Ghz}(Jy)$.   This predicts a thermal 5~GHz flux of 21~mJy. This is somewhat greater than the $14.5\pm0.3$ mJy  at 4.8~GHz   \citet{KO99} report from  high resolution maps 
but consistent with the 27~$\pm$ 3~mJy reported in lower resolution maps
 \citep{1990A&AS...82..273Z} 
 which suggests that neon is also near solar abundance.

\section{Ionized Gas Kinematics in He 2-10}

\subsection{ Localized Velocity Components}
\subsubsection{The Two Emission Regions are at Different Velocities}
Figure 2 shows the [NeII] spectra at every point in the mapped region, and Figure 3 shows velocity channel maps.  The two regions East (E) and West (W) are clearly offset in velocity: the W source appears between 915-870\kms, 
centered at $v_W\sim 891$~\kms, 
and the E between 890-840 \kms, 
centered at  $v_E\sim 864$~\kms. 
The fits to the lines are discussed in the next section. 
  \citet{AH07} found a similar overall offset in the Brackett emission of the two regions. 
\citet{SA09} and \citet{KO95}
found the molecular gas to be rotating  with the steepest velocity gradient at  $P.A.~130^{\circ}$, $\sim25^{\circ}$ east of the line joining the two infrared sources.  The relative velocities of the two [NeII] regions agree in sense and magnitude with the observed rotation; but the velocity at this spatial resolution does not show any signs of rotation, shear or streaming within each source. 

\subsubsection{The Non-thermal Source has no Kinematic Signature} 
It has been suggested \citep{RE11}  that the non-thermal radio/x-ray source, 
possibly an AGN, 
 between the two infrared lobes may drive an outflow.   If such an outflow impinged on or entrained the gas which produces [NeII], it might affect the kinematics detectably.  We do not find any evidence for outflow in the [NeII] spectra and line profiles in the area of the non-thermal source are consistent with the two emission regions overlapping in one beam.  If there is an outflow, it must be entirely in the plane of the sky 
 or not coupled at all to the low excitation gas of the star forming region.
 Observations with spatial resolution capable of cleanly separating the two regions could answer this question more definitely.  \nopagebreak

\subsection{The Integrated [Ne II] Profile}
We now look at the main component of the [NeII] line, which contains the bulk of the ionized gas.   To what extent is the line profile gaussian?  Gaussian line shapes are produced by random motions, which for these nebulae include thermal and gravitational turbulence.   Parameters of the turbulence are found from the best-fitting gaussian,  and the discrepancies between the fits and the observations show departures from gaussianity--i.e., non-random, bulk motions.   
So finding the best-fitting gaussian is an important step in understanding the gas motions.  

Unfortunately, there are significant systematic problems in fitting the [NeII] lines.  Velocities more blue than $\approx 860$\kms\ fall in a region of low grating efficiency and increased noise. 
 The red and blue sides of the lines appear different on inspection, as well as in fits.  Since the red side of the line has less noise and no obvious non-gaussian features (except for a narrow feature at $873$\kms\ , discussed below, which can be removed), while the blue side has the known grating problem,  we choose to rely on the line profile red of the peak.  By reflecting the red data across the line peak we produce simulated symmetric spectra.  In the next sections we discuss the best fits for both the original and the symmetrized profiles, and what they reveal about the gas kinematics.  

 The original data for the line averaged over the W region can be fit (Figure~4) by a single gaussian with FWHM 62 \kms\ and centroid 888\kms\ but the $\chi^2/d.o.f.$ is poor, 2.32. It is clear from the figure that the severe discrepancy between data and fit  occurs at velocities blue of $\sim860$\kms.   The symmetrized line profile has a much better fit with  FWHM  $55\pm1.1$\kms\ , centroid $ 891 \pm0.9$\kms\ 
 and $\chi^2/d.o.f.=1.2$ (Figure~4).   The original data, the symmetrized profile, and the fit do not differ significantly from each other at velocities redder than  $860$\kms\ but at bluer velocities the data significantly exceed the main gaussian fit. This forms the  `blue bump'  seen in the channel maps.   A two-gaussian fit, shown in the figure, finds that the blue bump is at 840\kms~ and has FWHM 30\kms,  but the formal significance of the fit is low and the results should be considered only indicative.  
 
  We conclude from the line profiles and fits that the line in the western source can be characterized by a single gaussian with centroid 891 \kms\ and   FWHM about 55 \kms\ and an additional  `blue bump' around 830 \kms.

The E region line profiles are more problematic than the W.  Since the regions's overall velocity is about 30\kms\ blue of the W region, the entire blue half of the line is at velocities badly affected by the grating efficiency. In addition, the line is wider than in the W, leaving less baseline.  So the fits to the eastern lines are formally less good, and the differences between the symmetrized and original line shapes are more substantial, than in the west.  The profile averages over the entire eastern source can be fit with one gaussian with FWHM $80.4\pm1.4$\kms~ centered at $864\pm0.9$ \kms; this gives $\chi^2/d.o.f=1.9$.  The fit and line are in Figure 5a and it may be seen that the fit matches the blue side of the line, rather than the red side which has lower noise.    If we force a fit to the red side by using the symmetrized profile (Figure 5b), it is significantly narrower, with FWHM  $62\pm1.5$\kms\ , and  $\chi^2/d.o..f.= 2.1$   The residuals show that the data exceed the fit for the entire blue side of the line.  

The increased blue noise and the blue excess in the $Br\alpha$ profiles argue that the red side of the line is a better reflection of the turbulent gas motions.  Another reason to prefer the fit to the red side is that the FWHM of the turbulent motion depends on the mass concentration $M/R$; the FWHM from the red side agrees better with other observations of these sources than does the blue (this is discussed more fully in the next section).  In short, we favor the interpretation that the line in the E region has a main component with FWHM around 62\kms~ and significant excess emission at velocities from $\sim770$\kms~ to $\sim860$kms~.    The blue excess is too weak and too noisy to be meaningfully fit.   

\subsection{Cluster Parameters Derived from Line Profiles}
 
He 2-10 is a famous nursery for super-star clusters.  Images of the starburst region in the optical \citep{JL00} and near-infrared \citep{RE11} have revealed more than 10 star clusters in the E emission region.   The clusters observed change with the wavelength of the observation, showing  that the apparent structure is highly dependent on  extinction and that clusters have formed at different depths in the clouds.  The agreement of the [NeII] line flux with that predicted from the radio argues that the [NeII] is not significantly obscured and should penetrate to the same depth and probe the same clusters as does the radio. The main difference between them is that the radio has much higher spatial resolution. We therefore refer to the 3.5 cm maps of \citet{RE11}  which had spatial resolution $0.\arcsec44\times0.\arcsec21$, as a guide to the structure likely to underly the [NeII].  \citet{RE11} find four sub-sources within one [NeII] beam size in the east and two in the west, with each of the sub-sources so large that it probably includes several smaller clusters.   

While we cannot identify individual clusters from the current [NeII] data, we can deduce something of their physical structure from the turbulent motions seen in the line shapes.  The mass concentration $(M/R)$ is proportional to $\sigma_v^2/G$, where   $\sigma_v=FWHM_g/2.35$ is the 1-D velocity dispersion due to gravitational turbulence. The constant of proportionality relation depends on the mass distribution \citep{MRW88} and ranges from  $(M/R)\approx(5 \sigma_v^2/G)$ for constant density $\rho(r)$  to $ (M/R)\approx(2 \sigma_v^2/G)$ for $\rho(r)\approx r^{-2.25}$.   The gaussian line widths should be the sums in quadrature of the thermal widening, the instrumental resolution, and turbulent gravitational motions. The thermal line width for neon is 
 $4.04- 5.7$ \kms~ for $T_e$ from $7.5\times10^3$ to $1.5\times10^4$, so almost all the observed linewidth is due to gravitational turbulence.  The derived mass-to-radius relationship is  $\approx 2.7\times10^5M_\odot/pc$ for the W source and $\approx 3\times10^5M_\odot/pc$ for the E.  The formal errors due to the uncertainty in line widths are only about 
 4\%; much smaller than the 50\% uncertainty which is inevitable with an unknown mass distribution. 
 
  These mass concentrations are very like the concentrations deduced for NGC 5253 and II Zw 40 \citep{BTL13,BLT12}. The 80.4\kms~ FWHM obtained by fitting the blue side of the E source would give $(M/R)\approx 5.5\times10^5M_\odot/pc$, which would be the highest known; we take this as another argument {\it against} that fit.)  
The radius in these mass-radii relations cannot be the entire 50~pc emission region, as that would give total masses $\approx1.5\times10^7M_\odot$, which are much too high.   \citet{JL00} and \citet{RE11} found from their near-infrared and radio images that the star clusters have typical radii $<3pc$ (unresolved).  The total mass of stars can be estimated by comparing  STARBURST99 cluster models to the $N_{lyc}$ deduced from the radio continuum.  To produce the observed $N_{lyc}$ from a cluster with a Salpeter-Kroupa mass function and $0.10~M_\odot~\leq~M\leq~50~M_\odot~$  needs $M_{total}=1.4 \times 10^6~M_\odot$ in the W region and $3.8\times10^6~M_\odot$ for the E.  For upper mass limit $35M_\odot$, the total masses increase to $2.2\times10^6 ~M_\odot$ and $6\times 10^6~M_\odot$.  

The kinematics of the [NeII] thus agree with the images that multiple clusters are present in each emission region.  
  
\subsection{ Ionized Gas Kinematics and Interactions with Molecular Clouds}
\subsubsection{Distribution and Kinematics of Molecular Clumps in the Star-Forming Region}
The [NeII] spectra, because of the heavy neon ion, can trace the ionized gas with
thermal linewidths comparable to the velocity dispersions of individual GMCs. With 
submillimeter interferometry  molecular gas can be measured with spatial resolutions
comparable to the [NeII] slit. With these data we can compare CO and [Ne II]
both spatially and kinematically, a much more powerful means of finding associated cloud structures than is  spatial projection alone. This gives an unprecedented 
picture of the relation of the ionized and molecular gas.

In Figure 6 we show the CO(2-1) channel maps of \citet{SA09} with the
[NeII] contours from TEXES superimposed.  The [NeII] has been convolved to 2\arcsec\ resolution and the velocity channels selected to match the CO, which has 1.9\arcsec\ $\times$ 1.3\arcsec\  beam and 5\kms\ channels. The CO velocities were reported in LSR, to which we have added 13.3\kms\  to transform them to heliocentric, consistent with the [NeII]. The CO channels are matched to the [NeII] channels as closely as possible given the different velocity sampling, and in all cases to within 3\kms~. There are many distinct CO(2-1) clumps in the Santangelo map; they list 14 clouds based 
on a spatial and kinematic cloud separation using Clumpfind \citep{1994ApJ...428..693W}.  Only a few of the clumps are associated with  bright [NeII] emission.
 
The clump most clearly associated with a [NeII] feature in both space and velocity is Cloud 12.  It is compact,  relatively isolated, and has heliocentric 
velocity of 908 \kms .  
velocity.   Cloud 12 is a GMC with
an derived virial mass of $3.1\times10^6 ~M_\odot$ and $M_{gas}$ from the CO intensity of  $1.9\times10^6 M_\odot$  \citep{SA09}.   Gas masses derived from CO lines are uncertain, especially when the clouds are large and include regions of different characteristics.  We start with the two assumptions that a) these dense, starforming clumps will not have significant  amounts of sub-thermal CO, and b) the virial mass based on the linewidth is a good upper limit on the total mass enclosed; velocities will not in any case be {\it lower} than gravitationally induced.   We also consider the mass of stars embedded in the cloud. This factor is often neglected but the stars will affect the CO linewidths and may influence the masses found from $X_{CO}$ as well.  The radio continuum in He 2-10 has shown that the stellar masses are on the same order as the gas mass.  If the clusters are embedded in the clouds, their masses will contribute significantly to the virial mass.  For the W source, if all the star formation is fueled by Cloud 12,  the star formation efficiency $\eta\equiv\frac{M_{stars}}{M_{stars}+M_{gas}}$ is $\frac{1.4\times10^6~M_\odot}{3.1\times10^6~M_\odot}\approx45\%$ if the clusters are embedded in the cloud. If the clusters are not inside the cloud and $M_{gas} \sim M_{virial}$,   the mass found from the CO intensity is more appropriate.  That gives  $1.9\times10^6 M_\odot$ for  $M_{gas}$ and $\eta\approx$42\%.   We estimate uncertainties in $\eta$ of $\pm20$\%, mostly from uncertainty in the stellar mass, but even so $\eta$ is very high compared to the 1-5\%  typical of Galactic GMCs \citep{MO14}. 

The brighter nebular emission of the eastern E source overlaps a collection of GMCs but is not clearly associated with any one cloud.  Most of the eastern [NeII] emission is 
located between GMCs 5 and 6 which have heliocentric velocities 
863 and 868 \kms\, respectively.   
This is
close to the 864 \kms\ we derive based on the integrated line profile for the E source,  
and also 
near the velocity 
of the ``red spike" discussed at more length in the next section. The  [NeII] emission in the E also
extends to very blue velocities, out to 823 \kms~ heliocentric; there is no component of molecular gas that blue. The fact that a) the eastern [NeII] source is located in between
Clouds 5 and 6 and b) that it extends far to the blue suggests the presence of an outflow 
directed toward the observer.  The  E  
region thus appears to be at a more 
evolved state than the more deeply embedded west, even though there is more molecular gas in the vicinity.   The star formation efficiency computed from all the GMCs in the E would be considerably lower than in the more isolated E source; however, because of the velocity information we are
able to associate the star formation with CO clouds 5 and 6 of \citet{SA09}. The star formation efficiency including  both and only Clouds 5 and 6 is  $\frac{3.6\times10^6M_\odot}{11\times10^6M_\odot}$ or $\approx33\pm20\%$  whether  $M_{vir}$ or $M_{stars}+M_{gas}$ is used.   If the star formation is confined to one of the clouds $\eta$ will accordingly be higher: 50\% for cloud 5 alone and an unphysical 92\% for cloud 6. That the E region appears between clumps 5 and 6 suggests that the young stars may have opened up a cavity in the cloud, and the very blue velocities support this picture.

There is substantial CO east of the star-forming region which  has no ionized gas or embedded star formation. The most eastern [Ne II] component is a weak blue feature at 828 \kms\  near Cloud 11 at 833 \kms~ heliocentric and Cloud 4, which is nearly coincident with Cloud 11 in space (0.2s separation) and at heliocentric velocity 863\kms .   The spatial coincidence and significant (30 \kms) kinematic shift, coinciding with a weak [NeII] feature, are intriguing; 
could this be star formation produced by a cloud collision?

\subsubsection{The $873$\kms\ `Red Spike' }
The [NeII] data show a narrow emission feature at  $873$ \kms, seen in the spectra as a clear spike red of the line peak in the E and in the channel maps as a strong enhancement of the emission at $873$ \kms\ over the whole galaxy except perhaps on the extreme north-west edge.  Figure 7a shows channel maps around   $873$ \kms\  at high velocity resolution;  excess emission appears in 7 velocity pixels, more than 2 resolution elements, confirming its reality, and it is strongest relative to the main line towards the E and SE. Figure 7b shows the [NeII] line profile on the SE edge of the [NeII] region, which includes both narrow and wide components, and just E of that edge where  $873$ \kms\ feature dominates. The best fit to the `spike'  has FWHM of 5.1\kms\  and $\chi^2/d.o.f.$ of 1.6.  This  is consistent with the instrumental resolution and the  thermal broadening of neon at $T_e \approx 10^4 K$ added in quadrature; there is no contribution from bulk motion or gravitational turbulence.   
 
  We believe this component arises in the interface of the ionized gas and ambient molecular cloud material, and that molecular clumps involved are  \citet{SA09}'s clumps 6 and 10.  Those dense regions overlap with the red spike in velocity and  are slightly south of the main radio-infrared emission, where the $\sim$873\kms~ [NeII] feature is the strongest and most distinct.    The probable mechanism is simple:  ionizing stellar radiation reaches the face of the molecular cloud and creates a layer of $Ne^+$.  When first ionized it has the velocity and the very narrow velocity dispersion of the molecular cloud;  eventually it will be expand and become entrained in the turbulent motion of the \HII~ region.
         
    Gas ionized by young stars has been observed to interact with ambient molecular material in Galactic star-forming regions such as  NGC 3503  \citep{DU14} and W49 \citep{GA13}.  In most cases, however,  there is velocity data for the molecular component but not the ionized.   The TEXES observations of W51 \citep{LA07}  did record velocities in ionized gas  but concerned a jet interacting with a  cloud,  a situation very different from that of He 2-10.   This is the first case we know of in which the interaction of an extended complex of  \HII~regions and an extended molecular cloud is observed with velocity information on both parties.  

\subsubsection{Bulk Outflows--A Blue Channel? }
\citet{AH07}  found weak broad blue wings on Brackett lines in some positions in He 2-10.  They attributed the wings to high-velocity motion in the nebulae, driven by the winds of the embedded stars. Since this should be intrinsically symmetric in velocity  \citet{AH07}  suggested that extinction prevents us from seeing the red side of the broad feature.   The [NeII] lines  do not show this broad pedestal because it is too weak:  the broad wing in the strongest location is about 15\% the line peak and the [NeII] observations do not have the dynamic range to see this clearly.    The [NeII] profiles instead show strong blue excess much nearer the line peak, which was not apparent in the Brackett data because of the inherently lower spectral resolution of hydrogen lines.  The blue excess is different in the two emission regions. The W emission region has the `blue bump' $\sim45$\kms\ blue of the line peak and about 1/3 the peak intensity.   The channel maps in Figure 8 cover this velocity region with high resolution. They show that the `blue bump'  emission appears in at least 12 velocity pixels or 4 resolution elements. Its position and extent agree with  the main velocity peak.   In the E region the excess blue emission appears quite differently; instead of a distinct `blue bump' there is weak excess (relative to a purely turbulent profile) over a velocity range of $\approx 70$ \kms.  

The excess blue emission covers velocities blue of $\approx850$\kms\ and as may be seen in Figure 6 it does not coincide with any of the molecular clumps--iin fact it almost perfectly {\it avoids} the optical clumps.  This suggests that the blue-shifted material flows though low-density openings in the ambient material.

\section{Discussion: High Efficiency Creation of Multiple Clusters in Giant Clouds}

He 2-10 is a very unusual galaxy; to our knowledge it may even be unique.  Dwarf galaxies \footnote {With the exception of tidal dwarfs formed in interactions of larger galaxies; they are often strongly enriched in metals and gas} usually have little molecular gas and significantly sub-solar metallicity (\citet{VA02},\citet{TA98}); He 2-10 has roughly solar metalicity and copious ($\approx10^7 M_\odot$, \citet{HI13})  molecular gas.  Dwarf starbursts usually form only one or two super-star clusters at a time and typically hold many clusters at a range of ages, but only one or two very in the very young embedded stage \citep{Da98}; in He 2-10 more than 10  embedded super-star clusters appear to have formed coevally.  He 2-10 is thought to be a merger product \citep{KO95} and in some respects it resembles the starbursts in large interacting galaxies rather than other starburst dwarfs. But the He 2-10 starburst is much smaller than the classic interacting galaxy starbursts, both in spatial extent (a few hundred pc rather than kpc) and in intensity, with $L_{(FIR)}\approx 3\times 10^9 L_\odot$ \citep{HI13}.   

It is more revealing to compare He 2-10 instead to the most intense Galactic star formation regions. For example, the well-studied W49, the most luminous Milky Way star formation region, is a giant molecular cloud of $1.1\times10^6 M_\odot$ and ~110 pc extent that has formed multiple star clusters, including Young Massive Clusters (YMCs) of typical masses a few $10^5 ~M_\odot$ \citep{GA13}.  He 2-10 in some respects is W49 writ large;  the two star-forming regions have about twice the diameter and 10 times the mass of the W49 cloud \citep{SA09}.   But the total mass of young stars in He 2-10 is more than 10 times that of W49. It is about 50 times greater, because star formation in He 2-10 is at least 5 times more efficient than in W49.    This fundamental difference between star birth in He 2-10 and in W49 is also seen in the $M/R$ values deduced in the previous section, which are at least a factor of 10 higher than in W49 or elsewhere in the Milky Way and are typical of modern globular clusters \citep{MO14}. 

The kinematic data on the molecular and ionized gas has permitted us to associate certain GMCs with the star formation regions and to estimate the star formation efficiency $\eta$. $\eta$ is the main factor that determines whether the cluster will be gravitationally bound and long-lived, like e.g. globulars,  or whether it will be unbound and disperse, like OB associations.   In the simplest picture $\eta$ must be greater than or equal to $0.5$ for the system to be bound; more sophisticated modelling indicates that bound clusters may form with $\eta$  as low as $0.2$ (\citet{PF08} and references therein).   In He 2-10 $\eta \approx0.4$, much greater than the $\eta \leq 0.05$ usually seen in the Galaxy \citep{MO14}. He 2-10, unlike the Milky Way, is probably creating bound massive clusters.  

 The high efficiency of star formation in He 2-10 may reflect the galaxy's history as a merger product: such interactions increase pressure in the ambient medium, so that star formation can persist longer and reach larger $\eta$.  He 2-10 has the additional advantage that it is a dwarf, without the strong gravitational shear forces \citep{WBZ10} that can disrupt star formation in disk galaxies.

 \section{Conclusions}
 
We present spectra of [NeII]12.8 micron emission toward the central star-forming complex in 
the dwarf galaxy He 2-10. We have a spatial -spectral cube with 1.2\arcsec\  spatial 
resolution and effective spectral resolution, including thermal broadening, of 4-5 \kms~,
comparable to molecular gas observations. The cube covers both the eastern (E) and 
western (W) components of the star formation region.  We find that: 
\begin{enumerate}

\item The spatial distribution and strength of the [NeII] emission are consistent with radio continuum maps
and a solar Ne abundance.

\item At the sensitivity level of these observations there is no evidence in the [Ne II] for an AGN in between the eastern and western star-forming
complexes. 

\item From the [Ne II] linewidths we derive mass scales of $M/R \approx~ 2.7\times10^5~M_\odot/pc$  for the W
source and $3\times10^5M_\odot/pc$ for the E source. When combined with the stellar mass
predicted by the Lyman continuum rates from radio continuum and mid-infrared fluxes, these
concentrations are consistent with pc-scale structures, that is, massive stellar clusters.
Most of the [Ne II] emission arises in these pc-scale structures, consistent with radio
continuum images.

\item  A giant molecular cloud \citep[GMC 12 of][]{SA09} is clearly associated, both spatially and kinematically, with the [Ne II] in the
W source.  We obtain a star formation efficiency for this region of $\eta\approx40\%$. 

\item The association of molecular gas and clusters is less clear in the brighter E
 [Ne II] source, which is located in between CO Clouds 5 and 6 of \citet{SA09}. 
An efficiency based on all the GMCs in the eastern 
region would be much lower than that in the isolated W source, but we can rule this out
based on lack of kinematic association. 
$\eta$ calculated with the two GMCs which bracket the [NeII] emission spatially and
spectrally is similar to that in the W at $\eta \sim 33$\%.  
 This region has a localized blue flow extending to at least 70\kms\ blue of the peak velocity.
 The E region appears to be more evolved than the 
W region. 

\item A narrow spectral feature in the [NeII] spectra, the `red spike' at 873~\kms,
 corresponds in space and velocity with GMCs 6 and 10 of \citet{SA09}.
 We believe it shows the interface of the ionized and molecular gas.

\item In addition to the localized blue flow associated with the E region, we see
a widespread blue flow associated with extended [Ne II] emission and anti-correlated with the GMCs,
and possibly associated with the blue flow seen in Brackett recombination
lines by \citet{AH07}. This suggests that the blue-shifted material flows through
low density openings in the gas.
 
\end{enumerate} 

 \acknowledgments  TEXES at the IRTF is supported by NSF AST-0607312 and AST-
0708074 to Matt Richter. JT gratefully acknowledges the support of NSF AST-1515570. This research has made use of the NASA\&IPAC Extragalactic
Database (NED), operated by the Jet Propulsion Laboratory, Caltech, under 
contract with NASA.

\onecolumn
\begin{figure}
\begin{center}
\includegraphics*[scale=0.4, angle=-90]{FIG1A.EPS}

\includegraphics*[scale=0.25]{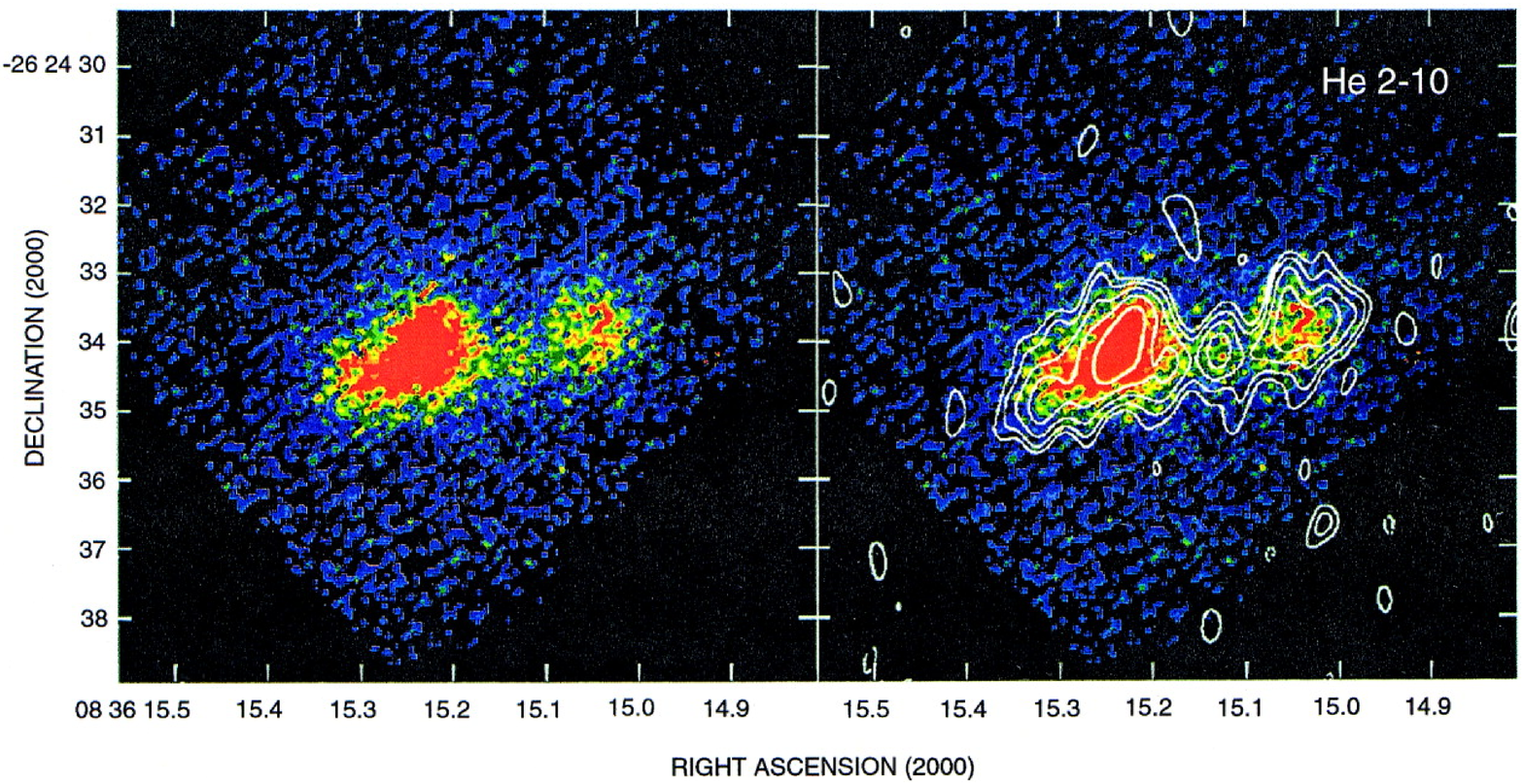}
\caption{Top: The spatial distribution of the [NeII] emission, found  by taking the zeroth moment of the data cube. The contours are integer multiples of $10\%$ of the peak value of $3.42 ~erg (s~cm^2~sr)^{-1}$. Bottom: The $12~\mu$m continuum emission (left) and the 2 cm radio contours (right) from \citet{BT01}.  The infrared spatial resolution was $0.3\arcsec \times 0.5\arcsec$. The radio contours are  $0.2~mJy~beam^{-1}$ times $\pm2^{n/2}$ and the beam $0.8\arcsec \times 0.4\arcsec$. }
\end{center}
\end{figure}

\begin{figure}
\begin{center}
\includegraphics*[scale=0.7]{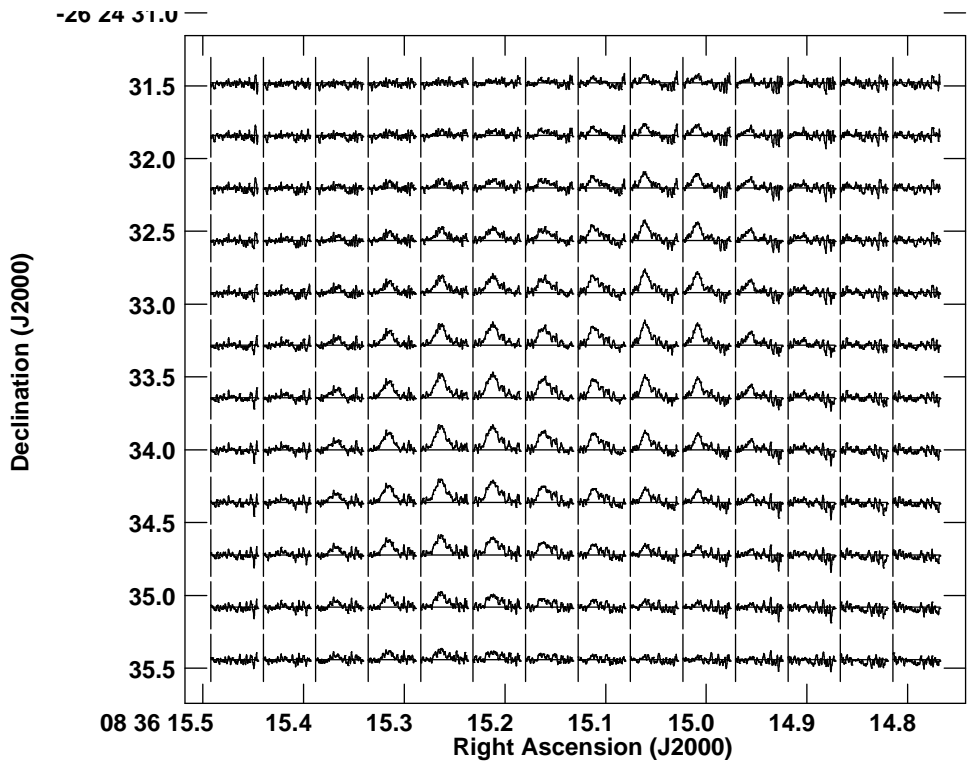}
\caption{The line profile at every position in the map. The data was convolved with a 1.4\arcsec~ beam and the continuum subtracted.  The absolute coordinates are estimated to be good to ~1\arcsec.    }
\end{center}
\end{figure} 

\begin{figure}
\begin{center}
\includegraphics*[scale=.9]{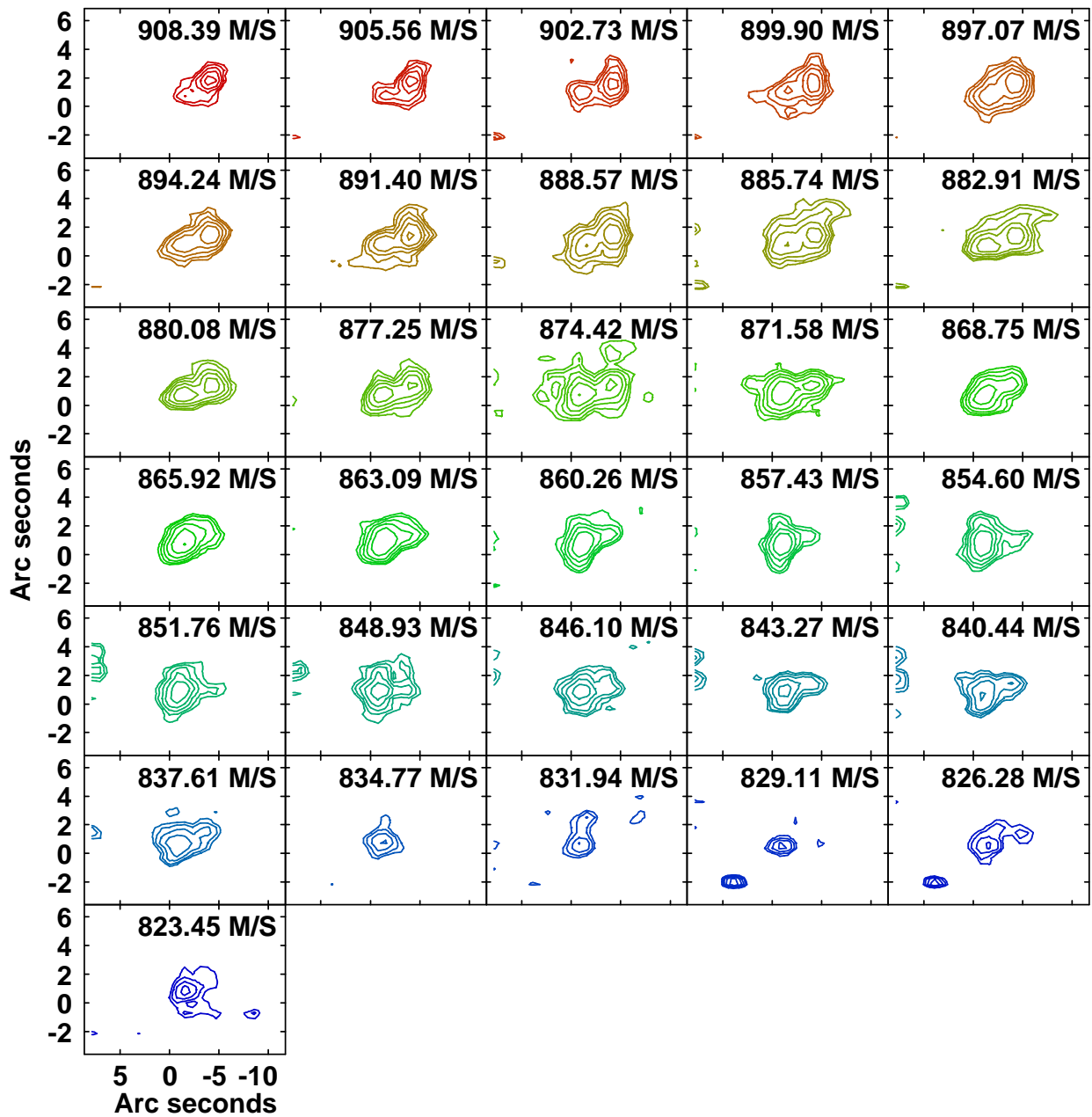}
\caption{Channel maps of the [NeII] line over the whole galaxy.  The panes are spaced by 3 pixels or one resolution element and are labeled by velocity.  Contours are $3,4,5...10\times 2.3\times10^{-3}~erg(s~cm^{-1}~cm^2~sr)^{-1}$. The colors of the [NeII] contours here and in subsequent figures are for eye relief only.} 
\end{center}
\end{figure}

\begin{figure}[h]
\begin{center}$
\begin{array}{cc}
\includegraphics[width=3.5in]{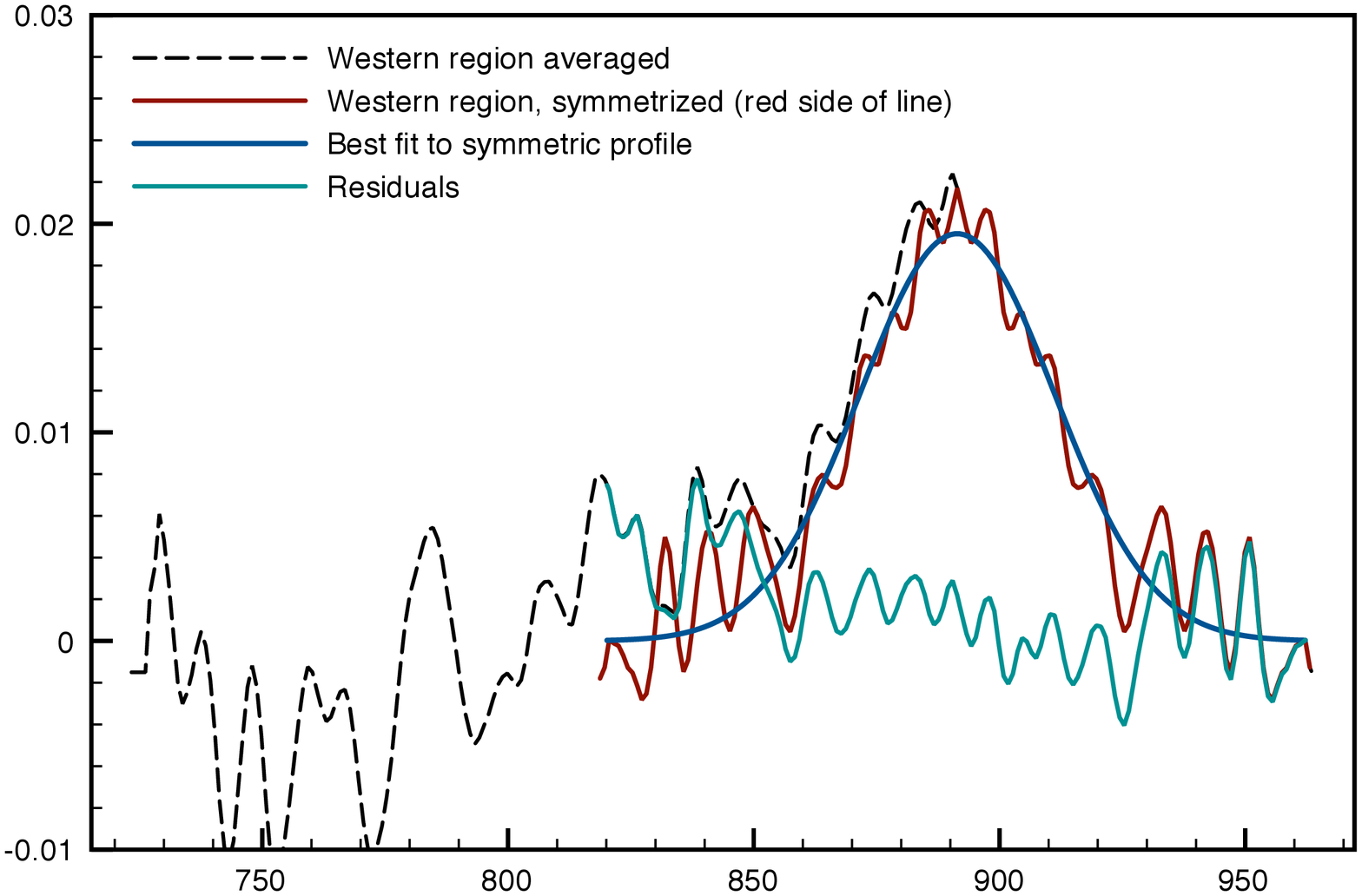} &
\includegraphics[width=3.5in]{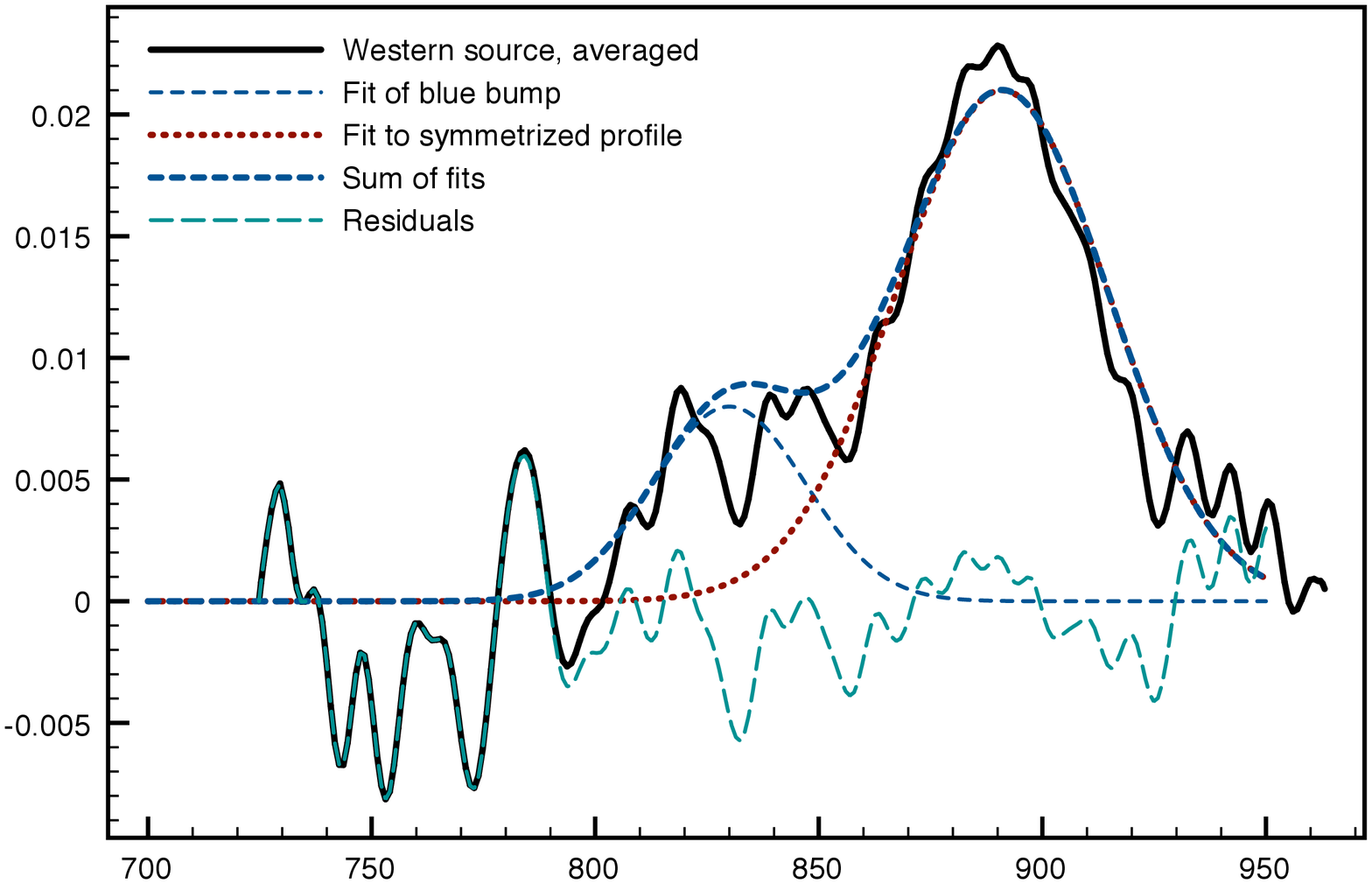}
\end{array}$
\caption{Left:  The [NeII] line averaged over the W source,  the symmetrized profile made by reflecting the red half of the line, the best single Gaussian fit to the symmetrized data, and the residuals of the original, non-symmetrized data.  X-axis units are  \kms~ and Y axis units are  $erg(s~cm^{-1}~cm^2~sr)^{-1}$. Note that the original data and the symmetrized are very similar except in the high noise region around 830 \kms.  Right: The same as the left panel, with another gaussian appropriate for the `blue bump' added.  The formal significance of the second gaussian is low because of the high noise. } 
\end{center}
\end{figure}

\begin{figure}
\begin{center}$
\begin{array}{cc}
\includegraphics*[width=3.5in]{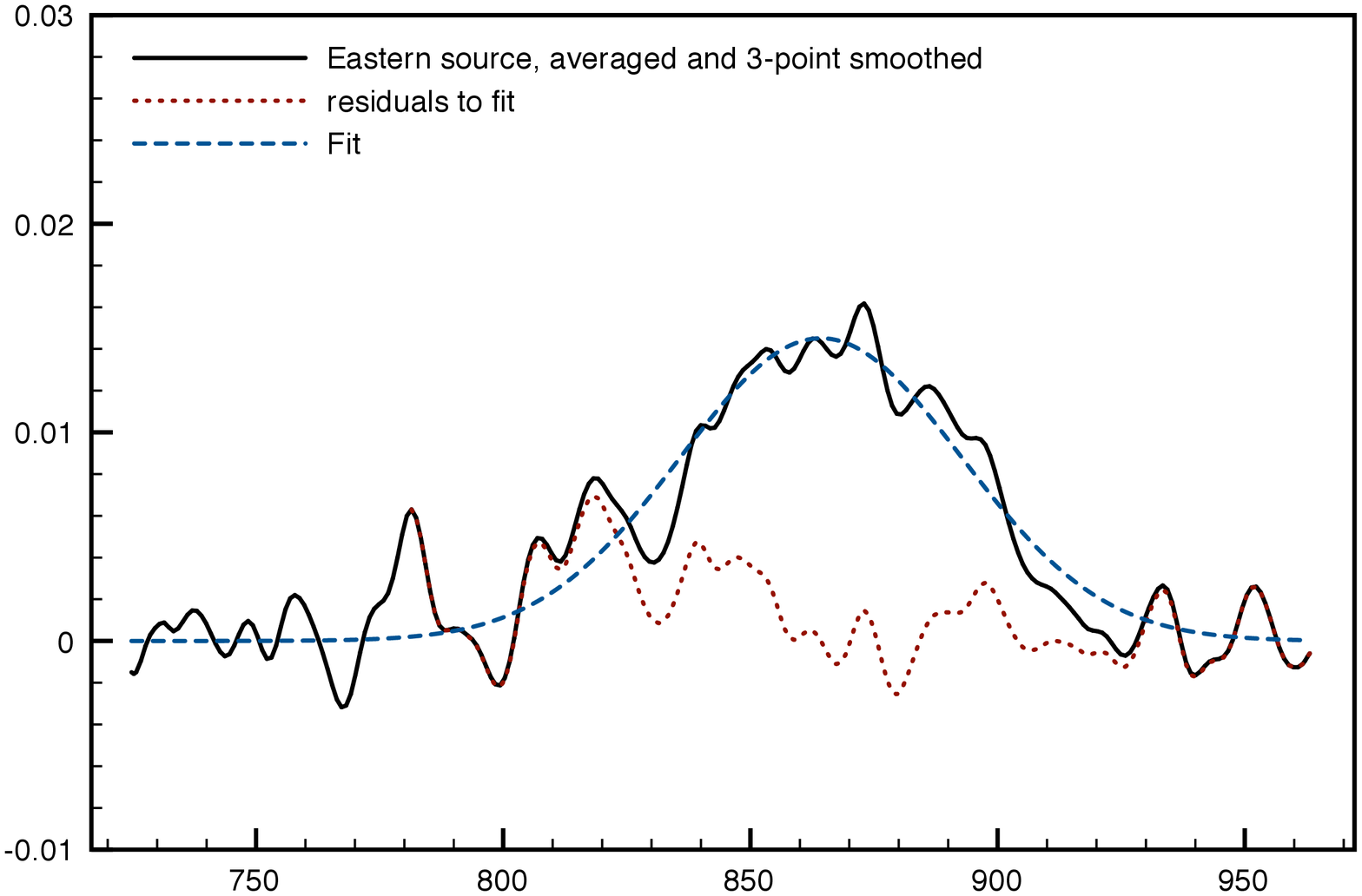} &
\includegraphics*[width=3.5in]{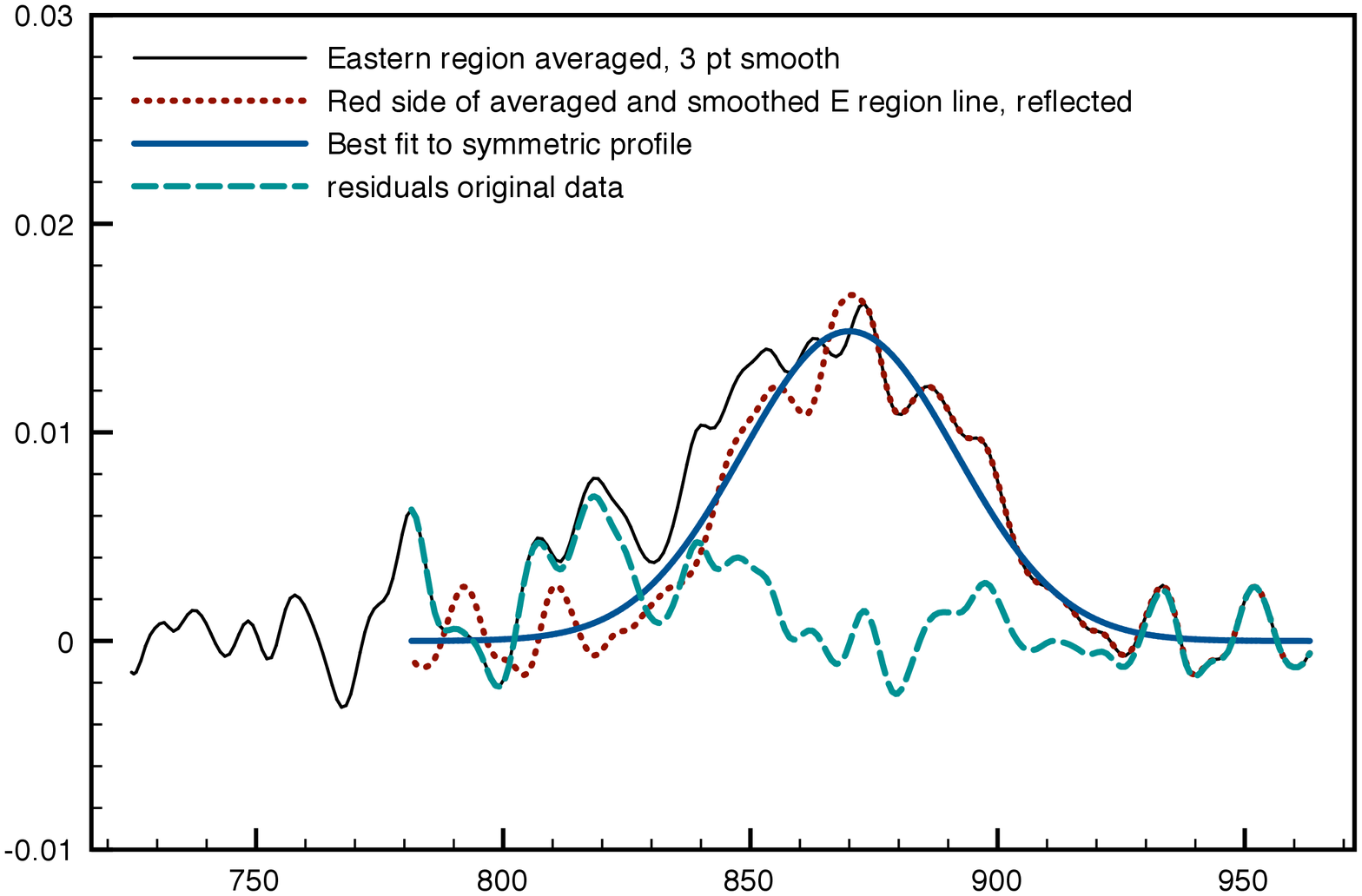}
\end{array}$

\caption{Left: the [NeII] line averaged and 3-point smoothed over the E source and the best single gaussian fit.  The fit is clearly better on the blue side of the line than on the left.  Right: The original line profile, the symmetrized line profile produced by reflecting the red half, the best single gaussian fit to the symmetrized data, and the residuals of the original data, for the E source.  X-axis units are  \kms~ and Y axis units are  $erg(s~cm^{-1}~cm^2~sr)^{-1}$. }
\end{center}
\end{figure}
\clearpage
\begin{figure}
\begin{center}
\includegraphics*[scale=0.7, angle=-90]{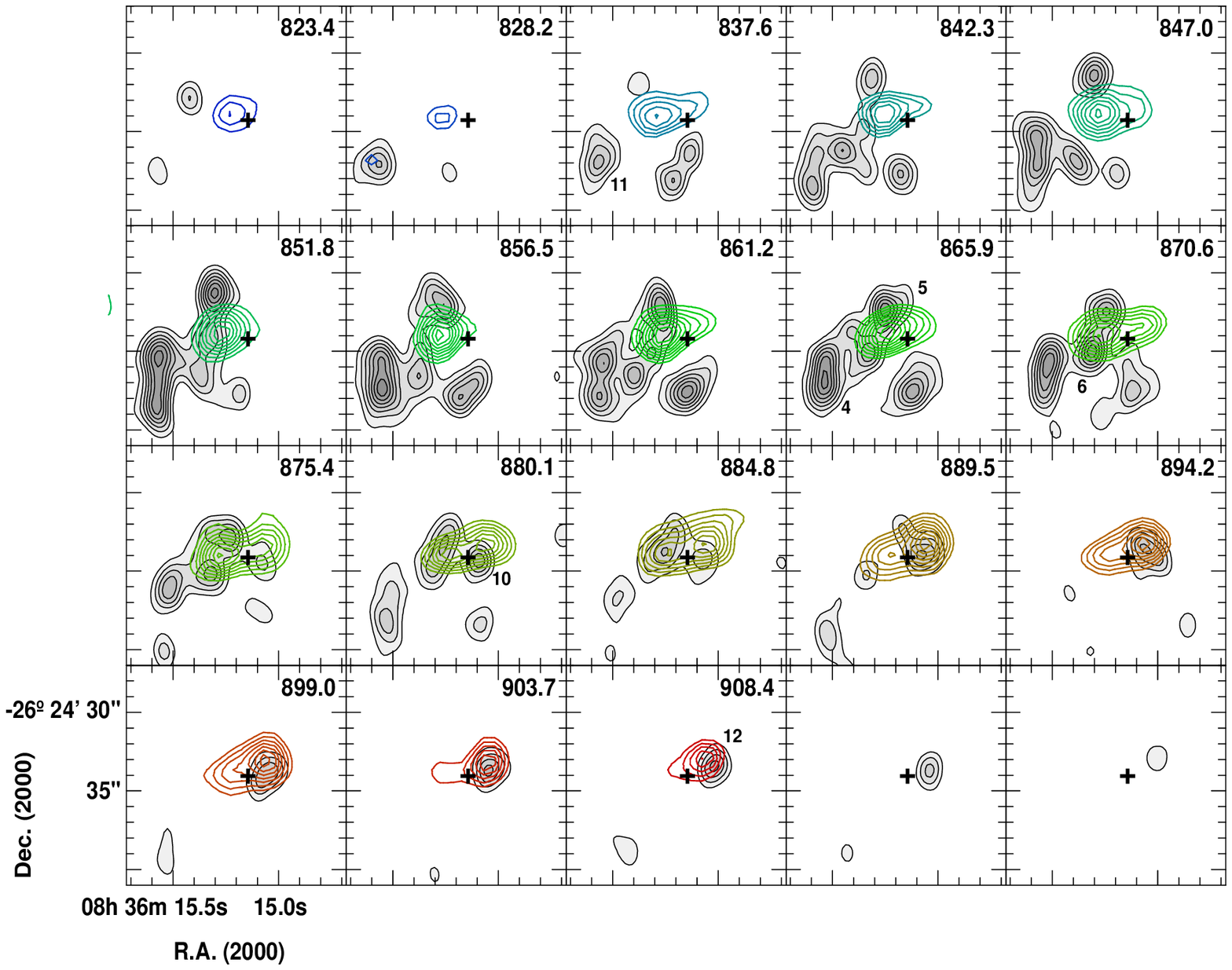}
\caption{The TEXES [NeII] emission in color overlaid on the CO(2-1) SMA channel maps of \citet{SA09}.   The panels are labeled by the [NeII] velocity channel.  The SMA channel velocities are within  3 \kms\  of the [NeII].  The molecular clouds discussed in the text are labeled with \citet{SA09}'s numbers. 
The + marks the position of the central nonthermal radio source, posited to be an AGN,
in between the eastern and western starforming regions (E and W in the text.) The
[NeII] spectra have been convolved to match the CO(2-1) 2\arcsec\ beam, and the 5\kms\  resolution 
of the CO maps.  }
\end{center}
\end{figure} 

\begin{figure}
\begin{center}
\includegraphics*[scale=0.5]{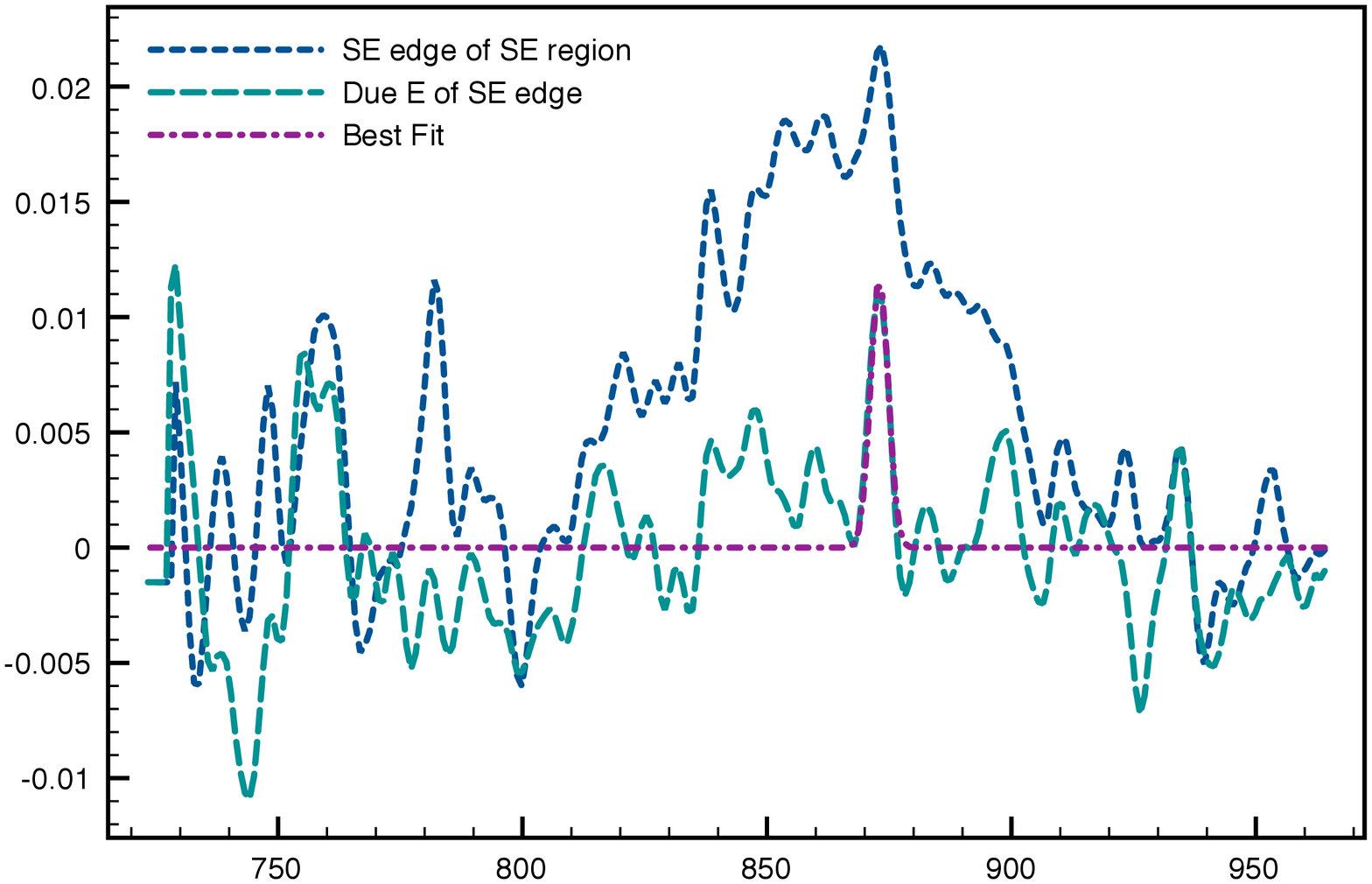}
\smallskip
\includegraphics*[scale=0.6, angle=-90]{FIG7B.EPS} 

\end{center}
\caption{Top: The [NeII] line profile on the E edge of the SE emission region and just off the emission region, with the best fit to the latter. X-axis units are  \kms~ and Y axis units are  $erg(s~cm^{-1}~cm^2~sr)^{-1}$. Bottom: Channel maps labeled by velocity over the velocity region of the 873 \kms~ feature. Contours are $2^{n/2}\times 7\times10^{-3}~erg(s~cm^{-1}~cm^2~sr)^{-1}$. The spacing between channels is one velocity pixel or $1/4$ of a velocity resolution element. }
\end{figure}

\begin{figure}
\begin{center}
\includegraphics*[scale=0.7, angle=-90]{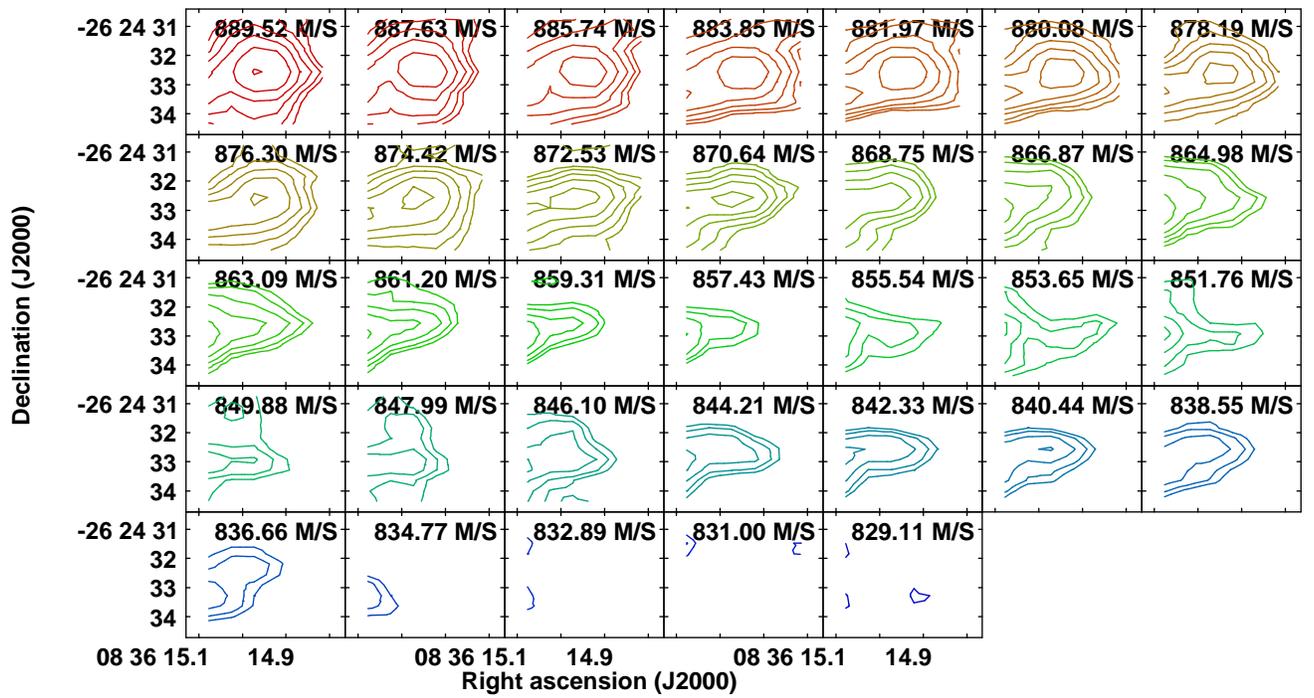}
\caption{Channel maps of the W emission region at velocities from the line peak through the `blue bump'. The panes are labeled by velocity and are spaced by 2 pixels, less than one resolution element.  Contours are $2^{n/2}\times7\times10^{-3}~erg(s~cm^{-1}~cm^2~sr)^{-1}$.  }
\end{center}
\end{figure}

\end{document}